\documentclass{aa}  
\usepackage{graphicx}
\usepackage{chemfig}
\usepackage[colorlinks=true,linkcolor=black,citecolor=blue]{hyperref}
\def\ts     {\thinspace}

\def\alphacii{\ifmmode{\alpha_\cii}\else{$\alpha_\cii$}\ts\fi}
\def\betacii{\ifmmode{\beta_\cii}\else{$\beta_\cii$}\ts\fi}
\def\gammacii{\ifmmode{\gamma_\cii}\else{$\gamma_\cii$}\ts\fi}
\def\deltacii{\ifmmode{\delta_\cii}\else{$\delta_\cii$}\ts\fi}
\def\cion   {\ifmmode{{\rm C}\scriptscriptstyle  \mathrm{II}}\else{C\ts {\scriptsize II}\ts}\fi}
\def\hei   {\ifmmode{{\rm He}\scriptscriptstyle  \mathrm{I}}\else{He\ts {\scriptsize I}\ts}\fi}
\def\heii   {\ifmmode{{\rm He}\scriptscriptstyle  \mathrm{II}}\else{He\ts {\scriptsize II}\ts}\fi}
\def\heiii   {\ifmmode{{\rm He}\scriptscriptstyle  \mathrm{III}}\else{He\ts {\scriptsize III}\ts}\fi}
\def\oi   {\ifmmode{{\rm O}\scriptscriptstyle  \mathrm{I}}\else{O\ts {\scriptsize I}\ts}\fi}
\def\oii   {\ifmmode{{\rm O}\scriptscriptstyle  \mathrm{II}}\else{O\ts {\scriptsize II}\ts}\fi}
\def\sii   {\ifmmode{{\rm ]Si}\scriptscriptstyle  \mathrm{I}}\else{Si\ts {\scriptsize I}\ts}\fi}
\def\siii   {\ifmmode{{\rm ]Si}\scriptscriptstyle  \mathrm{II}}\else{Si\ts {\scriptsize II}\ts}\fi}
\def\siiii   {\ifmmode{{\rm ]Si}\scriptscriptstyle  \mathrm{III}}\else{Si\ts {\scriptsize III}\ts}\fi}
\def\cristal{{\rm CRISTAL}}
\def\orchids{{\rm ORCHIDS}}
\def\yr{{\rm yr}}
\def\e  {\ifmmode{\rm e}\else{\rm e$^-$}\fi}%
\def\mstar {\ifmmode{\rm M_*}\else{\rm M$_*$}\fi}
\def\co  {\ifmmode{^{12}CO~}\else{$\rm ^{12}CO$~}\fi}
\def\ciii   {\ifmmode{{\mathrm{ C}}\scriptscriptstyle  \mathrm{I}}\else{C\ts {\scriptsize III}\ts}\fi}
\def\sfr{{\rm SFR}}
\def\cplus {\ifmmode{\rm C^+}\else{\rm C$^+$}\fi}%
\def\msun {\ifmmode{{\rm M}_{\odot}}\else{M$_{\odot}$\ts}\fi}
\def\lcii{\ifmmode{\mathrm{L_{\cii}}%
}\else{$\mathrm{L_{\cii}}$}\fi}%
\def\krome{{\sc krome}}
\def\gadget{{\sc gadget-3}}
\def\pgadget{{\sc p-gadget-3}}
\def\pgadgetk{{\sc p-gadget3-k}}
\def\iso{{\sc Iso}}
\def\corot{{\sc Corot}}
\def\counter{{\sc Counter}}
\def\perpendicular{{{\sc 90}-deg}}
\def\tdep{\ifmmode{\rm \tau_{\rm dep}}\else{\rm $\tau_{\rm dep}$}\fi}

\usepackage{amsmath}

\def\kms  {\ifmmode{{\rm \ts km\ts s}^{-1}}\else{\ts km\ts s$^{-1}$\ts}\fi}
\def\msol {\ifmmode{{\rm M}_{\odot}}\else{M$_{\odot}$\ts}\fi}
\def\lsun {\ifmmode{{\rm L}_{\odot}}\else{L$_{\odot}$\ts}\fi}
\def\cii  {\ifmmode{{\rm [C}{\rm \scriptstyle II}]}\else{[C\ts {\scriptsize II}]}\fi}
\def\ci   {\ifmmode{{\rm C}{\rm \scriptstyle I}}\else{C\ts {\scriptsize I}\ts}\fi}
\def\m    {\ifmmode{\mu {\rm m}}\else{$\mu$m}\fi}
\def\hi   {\ifmmode{{\rm H}{\rm \scriptstyle I}}\else{H\ts {\scriptsize I}}\fi}
\def\hii  {\ifmmode{{\rm H}{\rm \scriptstyle II}}\else{H\ts {\scriptsize II}}\fi}
\def\nii  {\ifmmode{{\rm [N}{\rm \scriptstyle II}]}\else{[N\ts {\scriptsize II}]\ts}\fi}
\def\oiii {\ifmmode{{\rm [O}{\rm \scriptstyle III}]}\else{[O\ts {\scriptsize III}]\ts}\fi}
\def\hh  {\ifmmode{{\rm H}_2}\else{H$_2$}\fi}
\def\nhh  {\ifmmode{N({\rm H}_2)}\else{$N$(H$_2$)\ts}\fi}
\def\microns {\ifmmode{\mu{\rm m}}\else{$\mu$m\ts}\fi}

\def\lya {\ifmmode{{\rm Ly}{\alpha}}\else{Ly$\alpha$\ts}\fi}
\def\ha   {\ifmmode{{\rm H}{\alpha}}\else{H$\alpha$\ts}\fi}
\def\hb   {\ifmmode{{\rm H}{\beta}}\else{H$\beta$\ts}\fi}
\def\ts     {\thinspace}
\def\cii  {\ifmmode{{\rm [C}{\rm \scriptstyle II}]}\else{[C\ts {\scriptsize II}]\ts}\fi}

\begin{document}

   \title{\cii as a global cold gas tracer in novel hydrodynamical simulations}

   \subtitle{}

   \author{Valentina P. Miranda
          \inst{1,2}
          \and
          Patricia B. Tissera \inst{1,2}
          \and 
          Jorge González-López \inst{1}
          \and
          Emanuel Sillero\inst{1,2}
          }

   \institute{Instituto de Astrofísica, Pontificia Universidad Católica de Chile, Av. Vicuña Mackenna 4860, 7820436, Santiago Chile\\
              \email{v.p.miranda@sms.ed.ac.uk}
         \and
             Centro de Astro-Ingeniería, Pontificia Universidad Católica de Chile, Av. Vicuña Mackenna 4860, 7820436, Santiago Chile.
            }

   \date{Received 28 May 2026, Accepted 28 September 2026}

  \abstract
   {\cii emission is a powerful probe of the gas content in galaxies, especially at early stages of evolution. Given its low excitation potential, the \cii 158 $\mu$m line is the main coolant of neutral gas in photodissociation regions and giant molecular clouds, and hence a tracer of all phases of the interstellar medium. However, there is no consensus on the nature of the gas traced by this emission.}
   {We aim to provide insights into the physics of the \cii emission and predictions for future observations.}
   {We used four pre-prepared simulations performed with a version of \pgadget, which includes the \krome\ chemistry package. We implemented a post-processing semi-analytical model to simulate the \cii emission. We analysed an isolated Milky Way-mass-size galaxy, and three merger configurations: a co-rotating, counter-rotating, and perpendicular major merger of two Milky Way-mass-size galaxies.}
   {We reproduce fundamental relations such as the \cii luminosity and the star formation rate, \lcii-SFR relation, in which mergers show an extra component for the \cii emission not traced by the star formation alone. During quiescent or starburst stages, galaxies deviate from the \lcii-SFR relation. We find that in our simulations \cii is a robust tracer of molecular gas, as well as atomic gas in both cold and warm phases on a global scale. Our results indicate that around 50 percent of the total \cii emission is traced by molecular gas, while 30 to 40 percent is traced by atomic and ionised hydrogen regardless of the initial configuration of the merger simulation.}
   {Our findings suggest that the nature of the \cii emission depends strongly on the merger history of galaxies, in which mergers and starbursts can act as more efficient drivers of \cii emission through collisions with molecular hydrogen.}


   \maketitle

\section{Introduction}\label{sec:introduction}

Unravelling how galaxies were assembled from small perturbations in the early Universe has been an inspiring but challenging task for astrophysicists \citep[e.g.][]{White&Rees1978}. Therefore, studying their star formation and chemical histories is fundamental to understanding how they formed and have evolved over cosmic time \citep{Tinsley1980,Madau1998}.

Cool gas in the Universe is one of the key ingredients for galaxy formation and growth, as it provides the immediate fuel for star formation \citep{Carilli2013}. The cold gas in the interstellar medium (ISM) is in the form of a cold neutral medium (CNM) \citep{McKee&Ostriker1977,Cox2005}.
Neutral hydrogen, \hi, is of particular interest to study as it is the main fuel for star formation. This is because \hi\ts can cool and transition into molecular gas, \hh, which is the gas from which stars are formed \citep{Carilli2013,Taconi2020}. Even though \hh\ts is the most common molecule in galaxies, tracing the cold gas in the form of molecular hydrogen is not trivial due to the challenging nature of \hh. This is mainly because \hh\ts lacks a permanent dipole moment and has a high excitation requirement even for the lowest transitions, with an upper energy of $h\nu/k \sim 510\ts \rm{K}$, which is much higher than the temperatures of giant molecular clouds \citep{Saslaw1967,Carilli2013}. Thus direct detection only traces a very small amount of the cold and dense gas, but not the bulk of it; consequently, other tracers are used to estimate the mass of molecular gas. Notably, recent direct detections of warm \hh\ts have been possible with the James Webb Space Telescope (JWST) \citep{Bialy2025,Kakkad2025}. 
 
Carbon monoxide, in the form of \co since it is the main isotope, is the one of the most used tracers of \hh\ts because it is the second most abundant molecule, and it has a low excitation requirement that is easily met by collisions with \hh\ts ($h\nu/k = 5.5$ $\rm K$ for the first state J = 1$\rightarrow$0  \co (1-0)) \citep{Carilli2013}. However, \co is not a reliable tracer of molecular gas in low-metallicity environments due to the low dust fraction that allows ultraviolet (UV) photons to penetrate deeper into the molecular cloud, photodissociating the \co molecules into a layer of emitting ionised carbon, \cplus. This is known as `CO-dark' gas \citep{Madden2020,Taconi2020}.
 
 The \cplus\ts ion can emit the \cii emission line in the wavelength $\lambda \sim$ 158 $\mu$m in the rest-frame. This line is produced by the forbidden transition $\mathrm{ ^2P_{3/2} \rightarrow ^2P_{1/2}}$ of singly ionised carbon, and it is a powerful tool to study the gas in the early Universe. Due to its low excitation potential (11.3 eV), it is the strongest emission line from cool gas with temperatures of $\lesssim$ 10$^4\ts\rm K$ residing in galaxies. It accounts for up to 1 percent of the total infrared luminosity \citep{Stacey1991,Helou2001,Diaz-santos2013}.

\begin{figure}
      \includegraphics[width=\hsize]{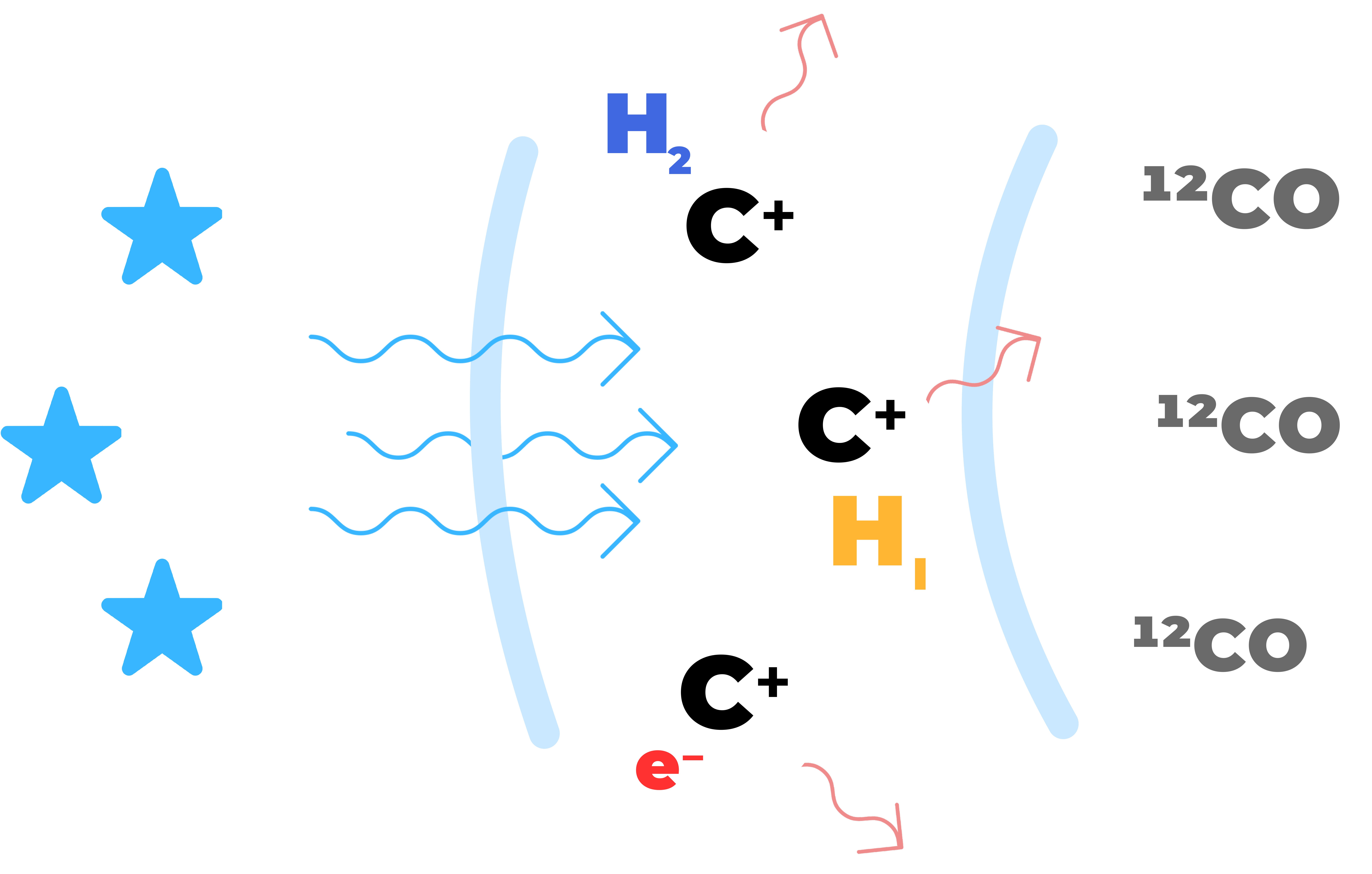}
      \caption{Scheme of \cii emission. Newly formed O/B stars emit UV photons (blue arrows) that dissociate \co, forming a layer of emitting \cplus. The \cplus\ts collides with \hh, \hi, and \hii\ts emitting \cii (red arrows).
              }
         \label{fig:scheme}
\end{figure}

The \cii emission line is produced by collisions of the carbon ion with electrons, atomic and molecular hydrogen \citep{Goldsmith2012}, as shown in Fig.~\ref{fig:scheme}. \cii is also the dominant gas coolant of neutral gas coming from photo-dissociation regions (PDRs). Hence, it is a robust tracer of all phases of the ISM with different density and temperature conditions \citep{Decarli&Diazsantos2025}, from ionised gas in HII regions to warm gas in PDRs \citep{Tielens1985}, and to cold molecular gas from gas clouds (GCs) \citep{Rodriguez-ambler2026}. Therefore, the interpretation of the gas phase traced by the \cii line becomes more challenging, particularly when spatially resolved observations are unavailable.

\citet{Stacey1991} estimated that about 70 percent of the total \cii emission from nearby spiral galaxies comes from PDRs. Other studies argue that 70 percent comes from molecular regions \citep{Olsen2015,Accurso2017}, or even 80 percent from neutral gas in star-forming galaxies \citep{Croxall2017}. \citet{Zanella2018} found a correlation of the \cii luminosity and the molecular gas content for ten main-sequence galaxies at $z\sim2$. This finding is also supported by other observational works, such as \citet{Madden2020}. Similarly, given that \cplus\ts has a lower ionisation potential than \hi\ts (11.3 eV $<$ 13.6 eV), neutral \hi\ts has been proposed as a dominant source of \cii emission \citep{Casavecchia2025}.

Additionally, \cii has also been proposed as a tracer of the star formation rate (SFR) of galaxies. This is because newly formed O and B stars produce far-UV photons that heat the gas on dust grains and polycyclic aromatic hydrocarbons via the photoelectric effect \citep{Helou2001}. Collisions of the \cplus\ts ion take place as the medium gets warmer, and if the gas is in thermal balance, these two factors, the SFR and \cii emission, as depicted in Fig.~\ref{fig:scheme}, are expected to be linked \citep{HerreraCamus2015}. Many works have estimated this relation for local galaxies \citep{HerreraCamus2015} and for diverse redshifts \citep{Delooze2014}. 

\citet{Schaerer2020} obtained a relation of $\log(\lcii/\lsun) = 6.61 + 1.17 \times \log(\sfr/ \msun \yr^{-1})$ for 118 star-forming galaxies at $z = 4.4-5.9$ from the ALMA Large Program to INvestigate C+ at Early Times survey \citep[ALPINE,][]{Faisst2020,Bethermin2020b}. The ALPINE survey is the largest \cii survey up to date, finding that around $\sim$ 40 percent of the sample showed signs of mergers, $\sim$\ts20 percent were dispersion-dominated galaxies, $\sim$\ts11 percent rotation-supported discs, and $\sim$\ts13 percent were compact systems \citep{LeFevre2020}. Hence, the \cii Resolved ISm STar-forming galaxies with ALMA \citep[CRISTAL,][]{Herrera-Camus2025} survey did a follow-up of 19 main-sequence galaxies at a higher resolution of $\sim 0.2"$. \cristal\ galaxies exhibited a wide range of behaviours, with well-behaved rotating gaseous discs indicating early disc assembly, a high frequency of mergers and interacting systems \citep{Herrera-Camus2025}, and the presence of galactic outflows \citep{Birkin2025}. Future efforts include the ORigin of the \cii Halos In Distant Systems \citep[\orchids,][]{Aravena2024} using JWST/NIRSpec, which aims to understand the nature of the extended \cii emission in star-forming galaxies at $z \sim 5-6$. 

In general, \cii emission has been observed in different types of galaxies \citep{Bradac2017,Harikane2018,Bakx2020}, up to very high redshifts \citep{Bouwens2022}, but there is still no clear consensus regarding the gas phase it traces. Several theoretical works have modelled \cii emission, including \citet{Ferrara2019}, who used analytical models, and \citet{Lagache2018,Popping2019}, who adopted semi-analytical approaches. Different implementations in cosmological simulations have been developed and tested using zoom-in simulations \citep[e.g. see][]{Vallini2015,Pallottini2017,Bibas2022,Schimek2024} or full cosmological runs \citep[e.g. see][]{Pallottini2022,Khatri2024,Munoz-Elgueta2024}. In particular, \citet{Casavecchia2024} uses \texttt{ColdSIM} hydrodynamical cosmological simulations \citep{Maio2022} to study \lcii\ts emission on galaxies at redshift $z \sim 6-12$. These works investigate how different properties of galaxies might influence the \lcii. \citet{Vallini2015} found that the \cii emission is dominated by PDRs. Similarly, \citet{Bibas2022} found that WNM is the primary source of \cii emission, as well as \hii\ts regions during active star formation. \citet{Schimek2024} reported that the \cii emission can be produced by the UV background and not stellar radiation. Their merger simulations show an extended component of \cii emission that is non-negligible ($\sim 10$ percent). This extended emission is produced by cold gas in the tidal tails between the main galaxy and the satellite.

Given that galaxy mergers can perturb the kinematics and modify the chemical signatures in interacting galaxies, studying how interactions influence \cii emission is of particular interest to interpret observations. Major mergers are characterised by the strong gas accretion towards the central regions of the system, induced by torques that redistribute the system’s mass and angular momentum \citep{Barnes1991,Tissera2000,DiMatteo2008}. \citet{Appleton2013} showed that \cii emission can be enhanced at large scales by turbulence and shocks, due to interactions in the Stephan's Quintet compact group. \citet{DiCesare2024} found large envelopes of diffuse gas emitting \cii around merging systems at high redshift. Some numerical works have studied how dwarf galaxy mergers affect the SFR of the interacting galaxies \citep{Bibas2022}.

However, there is no systematic study on how  mergers, and in particular major mergers,  influence the emission of \cii in galaxies, along and after the interactions and for different orbital configurations. Our simulations distinguish from previous works, in the main followin points. First, we follow the evolution of \cii and \cplus self-consistently along and after the coalescence.
Second, our chemical model is able to account for cold gas at temperatures lower than 100\ts K, and thus we can also separate the hydrogen gas in each phase, which allows us to effectively follow each component self-consistently over time.
Finally, we test different orbital configurations. To this end, we study the distribution of \cii emission and the abundances of \cplus, \hh, \hi, and \hii, and examine how these quantities evolve in relation to the star formation history of interacting galaxies and an isolated counterpart. We resort to pre-prepared simulations performed by using a modified version of the \pgadget\ code, which includes the chemistry package KROME, \pgadgetk, \citep{Sillero2021}. We aim to model \cii emission self-consistently to investigate its evolution during galaxy mergers, to constrain the physical origin of the emission, and to provide predictions for future observations. It should be noted that providing predictions of how our simulated \cii emission would be observed by ALMA is beyond the scope of this work and should include effects such as CMB supression, beam dilution, and sensitivity limitations. Hence, we provide predictions as a theoretical reference.

This paper is organised as follows. In Sect.~\ref{sec:simulations}, we provide a brief description of our simulations and the different merger configurations. In Sect.~\ref{sec:model} we describe the implemented model, as well as provide the validation and general trends. In Sect.~\ref{sec:global_relations}, we analyse general \cii relations and Sect.~\ref{sec:tracers} discusses the nature of the gas traced by the \cii emission. Finally, in Sect.~\ref{sec:conclusions} we summarise our main results and conclusions.

\section{Simulations} \label{sec:simulations}

In this work, we use a set of pre-prepared fully controlled simulations. In Sect.~\ref{sec:subgrid-physics} we briefly detail the implemented subgrid physics, and the main adopted assumptions. In Sect.~\ref{sec:simulated-scenarios} we describe the simulations for the isolated and merging scenarios, and present main trends.

\subsection{Subgrid physics}\label{sec:subgrid-physics}

The analysed simulations were performed by using \pgadgetk\ \citep{Sillero2021}, which is based on a modified version of the \gadget\ code \citep{Springel2005,Beck2016}. This version includes the chemistry package \krome\ \citep{Grassi2014}. The ISM is treated as a multiphase medium that allows coexistence and material exchange between the hot-diffuse gas phase and the cold-dense gas phase \citep{Scannapieco2006}.  
Stars are assumed to form from the cold-dense gas, for which we adopt an initial mass function (IMF) by~\citet{Salpeter1955}, with a lower and upper cut-off of 0.1 \msun, and 120 \msun, respectively. This choice was made following \citet{Nomoto2013}, who reported that Salpeter's IMF best represents the observations of the Milky Way (MW).

A fraction of these stars later explode as type II and type Ia supernovae (SNe), injecting energy and chemical elements into the ISM. The adopted model assumes that 70 percent of the released chemical elements and 50 percent of the SN energy are deposited in the surrounding cold phase, while the remaining chemical elements and energy are injected into the hot phase. The energy released into the hot phase is thermalised instantaneously, whereas the cold phase stores the injected energy in a reservoir until sufficient energy has accumulated to increase the gas entropy and promote the particle to the hot phase. This results in a self-regulated feedback process and naturally drives mass-loaded galactic winds whose strength depends on the depth of the galaxy's potential well \citep{Scannapieco2006,Miranda2026}.

The adopted chemical model follows the enrichment by SNe II, SNIa, and asymptotic giant branch (AGB) stars \citep{Mosconi2001, Jimenez2015, Tissera2025}. This model follows the enrichment of a set of 22 chemical isotopes, including \chemfig{^1H}, \chemfig{^4He}, \chemfig{^{12}C}, \chemfig{^{16}O}, \chemfig{^{24}Mg}, \chemfig{^{28} Si}, \chemfig{^{56}Fe}, \chemfig{^{14}N}, \chemfig{^{20}Ne}, \chemfig{^{32}S},\chemfig{^{40}Ca}, and \chemfig{^{62}Zn}.
Stars more massive than 8 \msun are assumed to end as SNII events with estimated lifetimes adopted from \citet{raiteri1996} and yields from \citet{Nomoto2013}. SNIa are assumed to originate from binary star systems, in which the primary star accretes matter from the companion star until it exceeds the Chandrasekhar limit. The yields for the SNIa are taken from \citet{iwamoto1999} with a delay time distribution (DTD) model for a single degenerate scenario \citep{Hoyle1960}. The yields for the AGB stars are an age-dependent function of the mass and metallicity, and are adopted from \citet{Karakas2010}.

In addition, as explained in detail in \citet{Sillero2021}, \pgadgetk\ considers the implementation of the code \krome, a chemistry package that follows the time-dependent evolution of different chemical species and gas temperature, given a determined set of chemical reactions or a chemical network, and solves this system of ordinary differential equations. In this implementation, the chemistry, and radiative cooling and heating are computed using \krome, following model 1a by \citet{Bovino2016}. This model considers photoheating, photoelectric heating, \hh\ts UV pumping, Compton cooling, atomic cooling, \hh\ts cooling, and chemical heating and cooling. 
It includes non-equilibrium photo-ionisational equations for nine primordial species: \hi, \hii, \hh, \hh$^+$, H$^-$, \hei, \heii, \heiii, and \e, and seven metal species: \ci, \cion, \oi, \oii, \sii, \siii, and \siiii. A linear system for the individual metal excitation levels for the most important coolants in the ISM is solved on the fly for a temperature of below $\rm T <10^4\ts K$, and are calculated using collisional partners from \citet{Wolfire2003,Richings2014}. Above this temperature, pre-computed photoionisation equilibrium (PIE) cooling tables from \citet{Shen2013} are considered. Overall, the chemical network covers 16 species and over 70 selected reactions. 

The \hh\ts recipe also considers formation through catalytic reactions on dust grain surfaces \citep{Jura1975}. Dust is taken into account as proportional to the metallicity of the gas particles following a dust-to-gas ratio 
of $D = D_\odot \cdot {\rm Z/Z_\odot}$ with $D_\odot = 0.00934$ \citep {Yamasawa2011} and $\rm Z_\odot = 0.0134$ \citep{Asplund2009}. Stellar particles represent a simple stellar population (SSP), adopting the models of \citet{Bruzual&charlot2003}. This includes the evolution of massive stars \citep{Chen2015} and spectral libraries to track their luminosity in ten energy bins ranging between 0.75 and 10000 eV. As described by \citet{Sillero2021}, the implemented subgrid physics model does not include a radiative transfer scheme, but instead adopts a phenomenological approximation, in which the incident radiation flux on a gas particle is then estimated by collecting the  contributions from stellar particles within a radius, $\rm R_c$, equal to the Jeans length. A dust attenuation model is also included, following a dust-to-gas ratio dependent scheme by \citet{Weingartner&draine2001}. Optical depth effects and self-absorption are also implemented and described in detail in \citet{Sillero2021} (Sects. 2.5 and 2.6). Finally, an uniform extragalactic UV background at redshift $z = 2$ is implemented  following \citet{Haardt&Madau2012}. This choice was made so that the simulations represent a galaxy evolving at cosmic noon, where the peak of SFR density is occurring \citep{Madau2014}.

The simulations have a spatial resolution of about 200 pc, and consequently clumps within molecular clouds (MCs), which have a typical size of $\sim 0.1-1$ pc \citep{Williams1994,Zhou2025}, cannot be resolved. To account for the molecular clouds' substructure, the adopted subgrid model implements an approach in which clumps are distributed within the MC according to a probability function dependent on a clumping factor, $\mathcal{C}_p$, which accounts for the possibly missed high-density regions due to numerical resolution. This approach is commonly used in simulations that aim both to study general properties and their evolution on galactic scales, and to understand the role of the physical processes acting on small and large scales \citep[][e.g]{Vallini2018}. In summary, although our simulations cannot resolve the very small scales within MCs, they were performed using a suitable scheme that accounts for this limitation \citep{Sillero2021}.

Finally, the star formation model follows a self-gravity criterion in which only cold and dense gas particles that cannot overcome gravitational collapse, will form stars \citep{Hopkins2018}. An important feature of this subgrid physics is that the SFR is coupled with the \hh\ts abundance, so that the probability of a gas particle being converted into stars is weighted by its \hh\ts fraction. This scheme has successfully reproduced the  Kennicutt–Schmidt law relation. Hence, within this scheme, the \hh\ts abundance is estimated self-consistently in the simulations, and the star formation activity is directly related to its abundance.
Additionally, the model self-consistently accounts for carbon production through three different stellar channels and follows the non-equilibrium evolution of the corresponding species. This provides an integral framework for studying the origin of the \cii emission.

\subsection{Simulated scenarios}\label{sec:simulated-scenarios}

We consider four scenarios: an isolated galaxy and three merger configurations, namely co-rotating coplanar, counter-rotating coplanar, and perpendicular mergers, hereafter denoted as the \iso, \corot, \counter, and \perpendicular\ts scenarios.
 All of them are constructed on the basis of the same initial conditions (ICs) for a single gas-rich disc galaxy. The isolated scenario serves the purpose of a `control case' to compare and determine how the galaxy's properties and the \cii emission evolve during interactions and mergers. 
 
 The galaxy's ICs are taken from \citet{Perez2011,Perez2013}, which contemplates a spiral MW-mass-sized (MW) galaxy with an initial stellar mass of $\mstar = 4.6 \times 10^{10}$\ts\msun and an initial gas mass of $\rm M_g = 3.57 \times 10^{10}$\ts\msun that conforms 40 percent of the disc component. The ICs account for a stellar bulge and disc, which, together with the initial gaseous disc, inhabit a dark matter halo in equilibrium.
 The initial mass of the gas particles is $m_{\rm g} \sim 1.25 \times 10^5\ts \msun$ with primordial abundances $\rm X_H$ = 0.75, $\rm Y_{H_{e}}$ = 0.25, and $\rm Z_{IC}=0$. Our MW isolated galaxy has a star formation history characterised by one main peak and a steady decline with time.

The \corot, \counter, and \perpendicular\ts mergers contemplate the interaction and coalescence of two MW galaxies with an initial separation of $\sim 280$\ts kpc and a relative velocity of 100\ts kms$^{-1}$. The \corot\ts and \counter\ts galaxies are located in a coplanar configuration; that is, the galactic discs lie in the same plane, but the latter galaxies rotate in opposite directions. In the \perpendicular\ts simulation, the galactic discs lie in perpendicular planes. 
We followed the evolution of all the configurations for 4\ts Gyr with a cadence of $\Delta t = 7 $\ts Myr through 602 snapshots available.
 
 The simulations are fully controlled and pre-prepared, i.e. the galaxies are already assembled at the beginning of the simulation and are not set in a cosmological context. The simulations have vacuum boundary conditions, in which the galaxies evolve in isolation with no external forces. This allows us to test particular merger configurations and helps us discern whether our results can be ascribed to the galaxy interaction itself.
 
\section{\cii emission model}\label{sec:model}

 As described in Sect.~\ref{sec:subgrid-physics}, \krome\ provides the abundances of \cplus, but since only ions of this specie that collide with \hh, \hi\ts, or \e\ts will emit as \cii, we implemented the emission model described by \citet{Casavecchia2025}.
 This model was tested in the {\sc ColdSIM} cosmological simulations \citep{Maio2007} by \citet{Casavecchia2024,Casavecchia2025},
 which has proven to predict the evolution of the cosmic mass density of the \cii emission, in agreement with observational estimations at $z = 6$ \citep{Davies2023} from high-resolution quasar spectra.
 
 We estimate the luminosity per gas particle as ${\rm L}_{\cii,p} = \Lambda_{\cii,p} \cdot V_p$, where the volume $V_{\rm p} = {\rm M}_{\rm p}/\rho_{\rm p}$ is given by the ratio of the particle's mass (${\rm M}_{\rm p}$) to the particle's density ($\rho_{\rm p}$), and $\Lambda_{\cii,p}$ is the power emitted per unit of volume. For the \cii emission line, which considers a two level transition, $\rm ^2P_{3/2} \xrightarrow{}^2P_{1/2}$ \citep{Hollenbach&McKee1989,Maio2007,Goldsmith2012,Casavecchia2025}, we can express
\begin{equation} \label{eq:lambda}
     \Lambda_{\cii} \equiv n_{\cplus,2} A_{21}  \Delta E_{21}
,\end{equation}
where $ n_{\cplus,2}$ is the number density of the \cplus\ts ions in the excited upper state, $A_{21}= 2.4\times10^{-6}\ts s^{-1}$ \citep{Draine2011} is the Einstein coefficient for spontaneous emission, and $\Delta E_{21} = 1.259 \times 10^{-14}{\rm\ts erg}$ \citep{Santoro2006} is the energy separation between levels. Following \citet{Maio2007}, we can rewrite Eq.~\ref{eq:lambda} in terms of the number density of the total \cplus\ts ($n_{\cplus}$)
\begin{equation}
\begin{split}
    &\Lambda_{\cii} = A_{21}\Delta E_{21} \cdot\\
    &\frac{n_{\cplus} (n_{\hh}\gamma^{\hh}_{12}+n_{\hi}\gamma^{\hi}_{12}+n_{\e}\gamma^{\e}_{12})}{n_{\hh}(\gamma^{\hh}_{12}+\gamma^{\hh}_{21})+n_{\hi}(\gamma^{\hi}_{12}+\gamma^{\hi}_{21})+n_{\e}(\gamma^{\e}_{12}+\gamma^{\e}_{21})+A_{21}}
\end{split}
,\end{equation}
where $\gamma^{\hh}_{12}$, $\gamma^{\hi}_{12}$, and $\gamma^{\e}_{12}$ are the excitation collisional rates, and $\gamma^{\hh}_{21}$, $\gamma^{\hi}_{21}$, and $\gamma^{\e}_{21}$ are the de-excitation collisional rates for \hh, \hi, and \e\ , respectively. We display the excitation collisional rates and constants used in Table~\ref{table:constants}, while we obtain the de-excitation collisional rates from the relation given by the detailed balance
\begin{equation}
    \gamma^{k}_{12} = (g_2/g_1) \gamma^{k}_{21} \exp{(-E_{21}/KT)},\ \ \ k = \hh, \hi, \e^-  
,\end{equation}
where $g_1$, $g_2$ are the statistical weights, which for the $\rm J_1 = 1/2\xrightarrow{} J_2 = 3/2$ jump corresponds to $g_1=2$ and $g_2 = 4$, and the excitation temperature is $\rm T_{exc}=E_{21}/k=91.2\ts K$.

We consider that for temperatures of $\rm T > T_h = 4\times10^4\ts K$ most carbon atoms are in higher ionisation states, such as \ciii. Therefore, we only take into account gas particles with $\rm T<T_h$, which allows us to correct for the potential overestimation of \cion. This assumption is widely adopted in other studieds of  \cii using numerical simulations \citep{Ferrara2019,DiCesare2024,Casavecchia2024,Casavecchia2025}.

\begin{figure*}
\sidecaption
  \includegraphics[width=12cm]{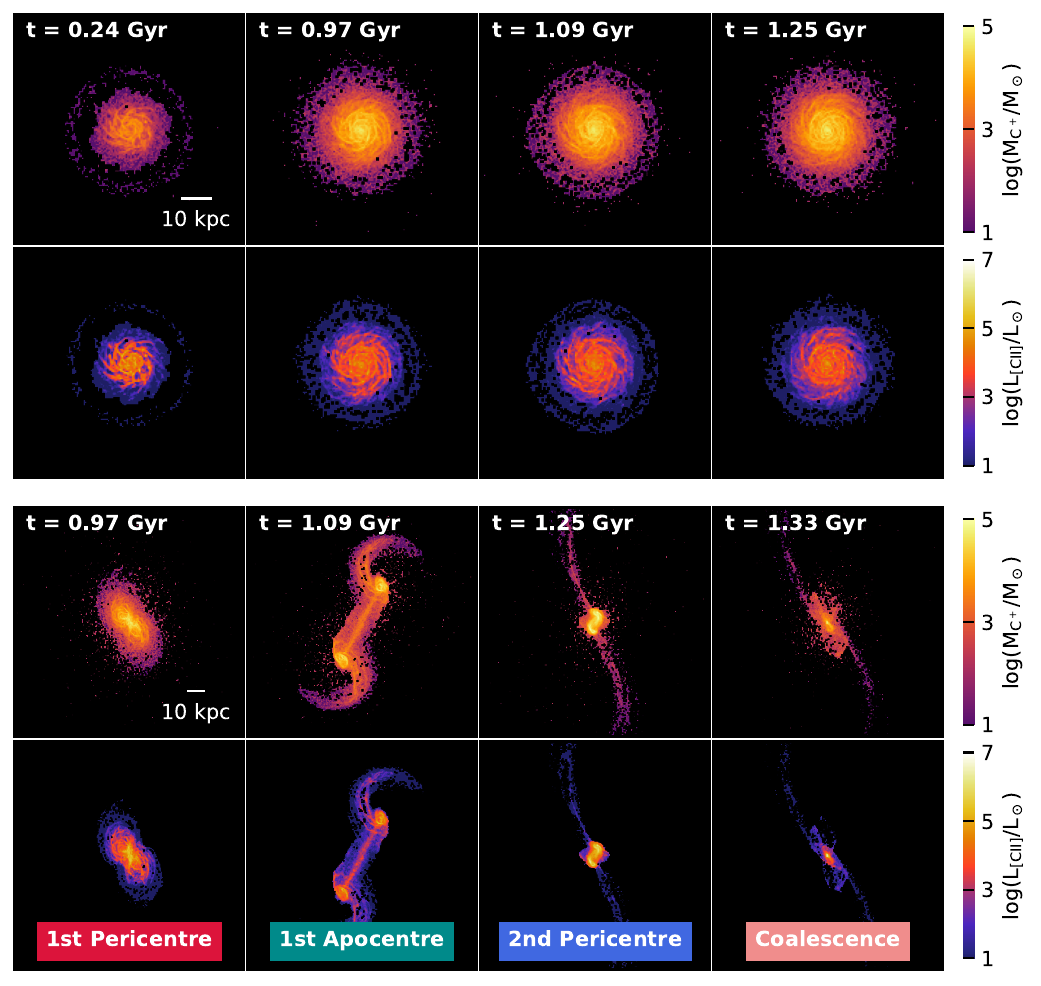}
     \caption{Face-on projections maps of ionised carbon mass (upper row) and of \lcii\ts (bottom row) with a pixel size of 0.5 kpc. Each column shows a specific time in the evolution of the \iso\ts galaxy (upper panel) and the \corot\ts simulation (bottom panel).}
     \label{fig:maps}
\end{figure*}

This model allows us to obtain the \cii luminosity for the gas particles according to their thermodynamical properties. As an example, in Fig.~\ref{fig:maps}, face-on projection maps of the ionised carbon mass (upper rows) and of the \cii luminosity, \lcii\ts (bottom rows) at different stages of the evolution of the \iso\ts (upper panel) and \corot\ts (bottom panel) simulations are displayed. These projections along the z axis were done considering a pixel size of 0.5 kpc to be consistent with the spatial resolution of the simulations. As seen, the \lcii\ts traces the spiral structure and arms in the \iso\ts experiment, pointing to a connection between star formation and \lcii, given that these are zones of active star formation. In the mergers, the \lcii\ts mainly traces the central parts of the galaxy, where star formation is occurring, as well as the stripped gas from the central parts of the galaxies, which forms a bridge between the galaxy pair.

\begin{figure}
    \centering 
    \includegraphics[width=\linewidth]{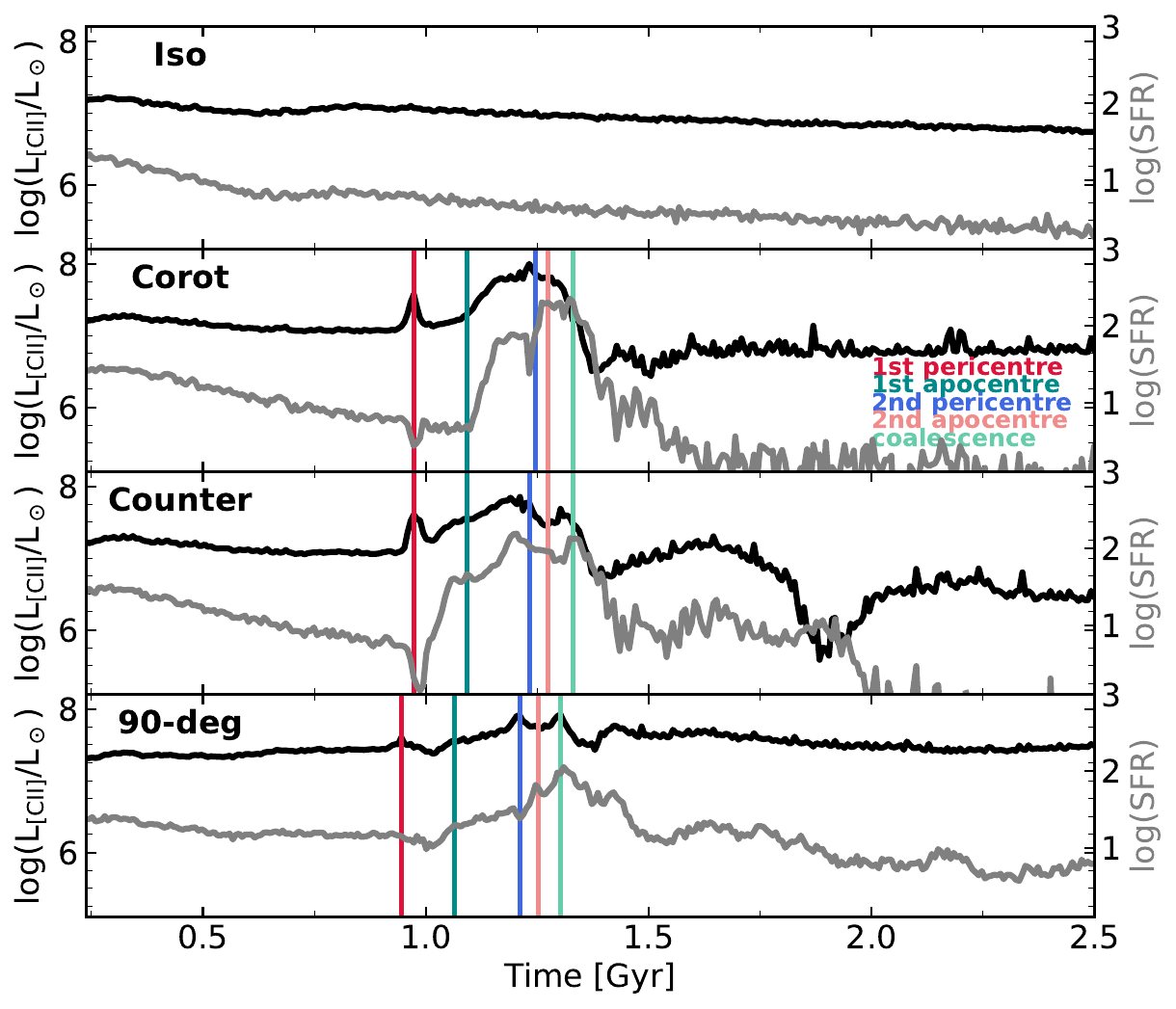}
    \caption{\lcii\ts (black lines) and SFR (grey lines) evolution in time. Each row shows the evolution of these quantities for our tested scenarios: \iso, \corot, \counter, and \perpendicular\ts mergers. For the mergers, key moments are shown in coloured lines in the bottom panel.}
    \label{fig:LCII_evolution}
\end{figure}

 We study the time evolution of the \lcii\ts in Fig.~\ref{fig:LCII_evolution}. The \lcii\ts in the isolated galaxy peaks with SF to then slowly decrease over time, indicating that both SF and \lcii\ts are correlated. The picture is more complex for the merger scenarios, in which a peak in luminosity is produced at the galaxies' closest proximity (pericentres), and it decreases as the galaxies move away from each other until the apocentres. During the first percentre, for both \corot\ts and \counter\ts mergers, a peak in \lcii\ts is accompanied by a decrease in SFR. This points to emitting gas that is not directly linked to SF, but rather to gas probably shocked during the interaction. In a follow-up paper, we study in more detail the nature of this emitting gas. Interestingly, after the merger and secondary burst of SF, the \lcii\ts stays constant through the evolution of the merged galaxy in the three merger experiments.
This constant emission remains approximately stable for the rest of the integration, suggesting that the system reaches a collisional balance state (see appendix~\ref{ap:bridge_cii}).
 
\subsection{Model validation}
To study how the \cii emission varies in each case during the galaxy-galaxy interactions, we first validated the implemented model to test how well fundamental relations are reproduced. For this purpose, we first used the \iso\ts simulation, as it is the simplest case, to then analyse how the mergers affect the expected relations.

As described in Sect.~\ref{sec:introduction}, a correlation between the \lcii\ts and the SFR of galaxies is widely observed for different types of galaxies, as well as at different redshifts. Thus, in Fig.~\ref{fig:lcii_sfr} we show the \lcii-SFR relation for each simulation, colour-coded by time. Each circle represents the state of a system at a given time of evolution. Observational relations, $\log(\lcii/\lsun) = 6.92 + 0.99 \times \log(\sfr/\msun \yr^{-1})$ from \citet[][solid black line]{Delooze2014}, $\log(\lcii/\lsun) = 41.24 + 0.97 \times \log(\sfr /{\rm erg} \hspace{0.1cm}{\rm s}^{-1})$ for local galaxies from \citet[][dashed black line]{HerreraCamus2015}, and $\log(\lcii/\lsun) = 6.61 + 1.17 \times \log(\sfr/\msun \yr^{-1})$ for star-forming galaxies at redshift $4.4 \leq z \leq 5.9$ from \citet[][dotted black line]{Schaerer2020} are shown for comparison. We note that we follow the simulated galaxies along their evolutionary paths, whereas observations report different galaxies at different stages of evolution. Hence, a comparison between simulated and observations should be done with caution.

As the system evolves, the isolated galaxy (upper left panel) moves onto the observed relations, globally in good agreement with the trend reported by \citet{HerreraCamus2015} and \citet{Schaerer2020}. The \iso\ts case shows a clear correlation between \lcii\ts and SFR, with a Spearman correlation coefficient of 0.98 (p $< 0.001$), and it evolves towards the observed relation over time.
However, a comparison should be done with caution as we discuss in Sect.~\ref{sec:tracers}, the \iso\ts is too simple to reproduce observational trends in all its complexity.

\subsection{Global \lcii\ts relations} \label{sec:global_relations}

\begin{figure*}
    \centering
    \includegraphics[width=\linewidth]{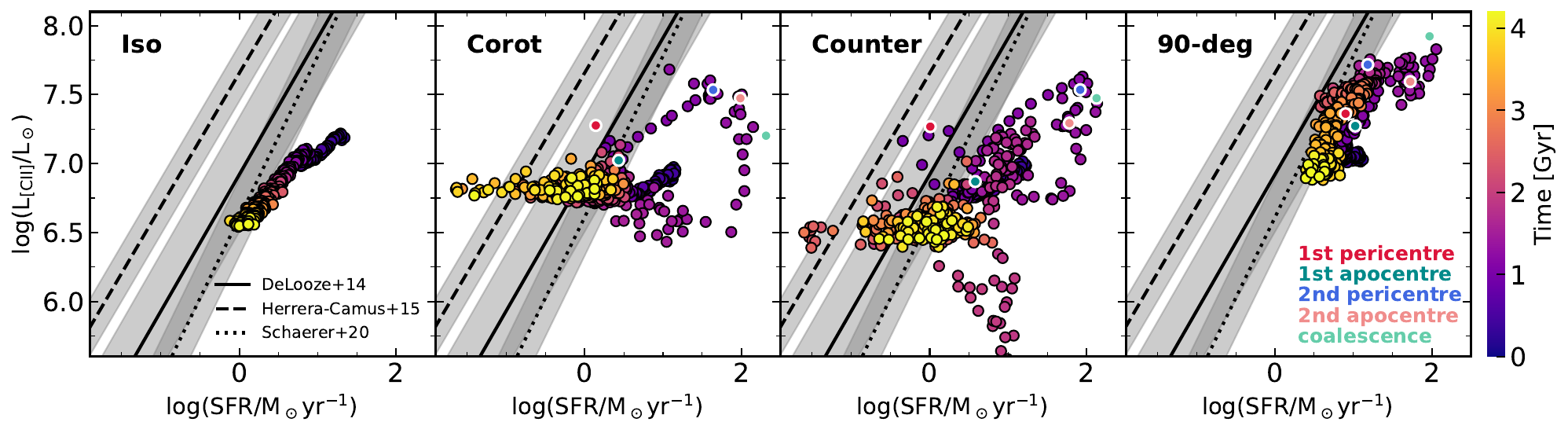}
    \caption{\lcii-SFR relation for the tested scenarios. Each circle represents the galaxy at a determined moment of its evolution (colour bar). For the mergers, a single galaxy is depicted until the coalescence (light green circle) happens. After the galaxies merge, the entire system is depicted as a whole. Each coloured circle represents the galaxy at a given merger time (inset labels). Observational relations are also shown for the entire sample of \citet[][solid black line]{Delooze2014}, local galaxies from \citet[][dashed black line]{HerreraCamus2015}, and star-forming galaxies at high redshift $4.4 \leq z \leq 5.9$ from \citet[][dotted black line]{Schaerer2020}, for comparison.}
    \label{fig:lcii_sfr}
\end{figure*}

We explore how the interactions impact the global \cii emission observed relations in Fig.~\ref{fig:lcii_sfr}. As expected, the mergers present a more complex behaviour, particularly the \counter\ts merger, given the nature of this configuration. For this reason, to facilitate comprehension, we have added in coloured circles to represent the merger times in the \lcii-SFR plane (see inset label). During the merger events, before the coalescence, we consider only the particles associated to the corresponding host galaxy to estimate \lcii\ts and SFR. After the coalescence, as both galaxies are merged, we estimate the properties considering all the system's particles with no distinction on their previous association to the progenitor galaxies.

Although mergers exhibit more chaotic behaviour, the galaxies follow a positive correlation with SFR, indicating that the star formation in these galaxies correlates with \lcii. Notably, the mergers show a loop behaviour, whereby an increase in the SFR produces an increase in \lcii\ts until the second pericentre (blue circle). After this, the galaxies get quenched and their SF ceases, which translates into lower a \lcii, showing a loop. For the \corot\ts and \counter\ts cases, the merged galaxy settles into a constant \lcii; that is, for a diversity of SFR ($\log{(\rm SFR/M_\odot)} = [-2,0]$), constant \lcii\ts is observed, which shows as a cluster of circles around $\log{(\lcii/\rm L_\odot)}\sim$ 6.5-7 (see upper right and bottom panels). This points to a decoupling between the SFR and the \lcii, in which additional physical processes might also be contributing to the \cii emission on a global scale in our simulations. This is particularly interesting as the \lcii-SFR relation is usually estimated using star-forming galaxies. 

The \perpendicular\ts case reproduces the observed correlation across all of its evolution. In this case, the \lcii\ts remains higher than in the \iso\ts experiment,  around $\log{(\lcii/\rm L_\odot)}\sim$ 7.5-8, even though the SFR range is similar. Interestingly, the \perpendicular\ts merger does not settle into a constant \lcii. This is because, as is seen in Fig.~\ref{fig:LCII_evolution}, after the collision, the \perpendicular\ts merger continues to actively form stars. Thus, when our simulated galaxies are forming stars, the primary source of \cii emission is star formation. In contrast, for the cases in which the resulting simulated galaxy gets quenched by the merger, the remaining constant \cii emission is produced by the past interaction. In Sect.~\ref{sec:tracers} we discuss the nature of this remaining emitting gas. Our simulations follow the expected relation and overall trend, with a median deviation for SFR of $\log(\textrm{SFR}/\msun\textrm{yr}^{-1})\sim$1, and with respect to the relation from \citet{Schaerer2020}, of -0.7\ts dex, -0.9\ts dex, -0.9\ts dex, and -0.7\ts dex for the \iso, \corot, \counter, and \perpendicular\ts simulations, respectively.

\begin{figure*}
    \centering
    \includegraphics[width=\linewidth]{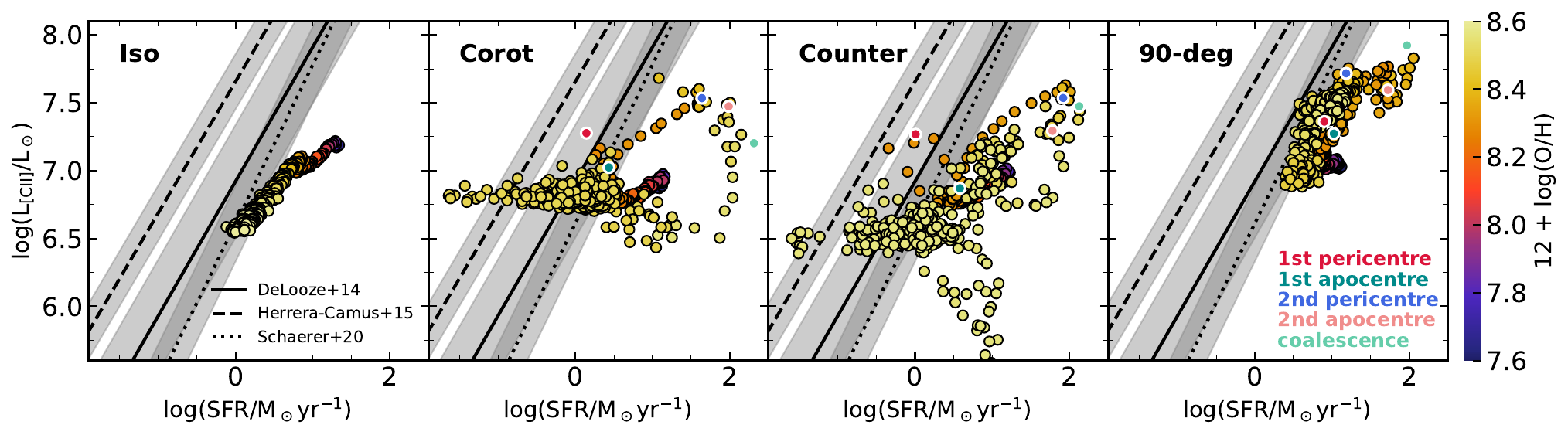}
    \caption{\lcii-SFR relation for the tested scenarios coloured-coded by the oxygen abundance of the galaxy's gas phase. Markers and observed relations are described in Fig.~\ref{fig:lcii_sfr}.}
    \label{fig:lcii_sfr_OH}
\end{figure*}

As we previously mentioned, our tested scenarios reproduce the observed \lcii-SFR relation, with the best agreement with the relation of \citet{Schaerer2020} for high-redshift galaxies. Given that at a higher redshift we expect a lower metal content in the overall gas phase, we explore a correlation between the \lcii-SFR relation and the gas oxygen abundance (coloured circles) in Fig.~\ref{fig:lcii_sfr_OH}. We estimate the oxygen abundance considering the total oxygen mass in the gas phase of the studied system, $\rm M_O$, relative to the total hydrogen mass in the gas phase, $\rm M_H$, i.e. $12 + \log(\rm O/H) = 12 + \log(\rm M_O/16) - \log(\rm M_H)$.

As can be seen from Fig.~\ref{fig:lcii_sfr_OH}, the oxygen abundance evolves accordingly with time (see Fig.~\ref{fig:lcii_sfr}) given that, as the galaxies evolve, they form stars that will enrich the gas phase with metals. The plateau formed around $\log{(\lcii/\rm L_\odot)}\sim$ 6.5-7 in the \corot\ts and \counter\ts mergers does not show a particular dependence on the level of abundance of oxygen, but rather has a constant value. In the \iso\ts case, we see that higher oxygen abundances lie at lower \lcii\ts and SFR values. The \perpendicular\ts case shows a weak dependence on the level of oxygen enrichment, whereby at a given SFR, higher \lcii\ts translates into higher oxygen abundances. This is consistent with previous observational and numerical works \citep[see][]{Delooze2014,Vallini2015,Lagache2018}. We discuss in Sect.~\ref{sec:tracers} how \cii and \hi\ts behave with metallicity in more detail.

\begin{figure*}
    \centering
    \includegraphics[width=\linewidth]{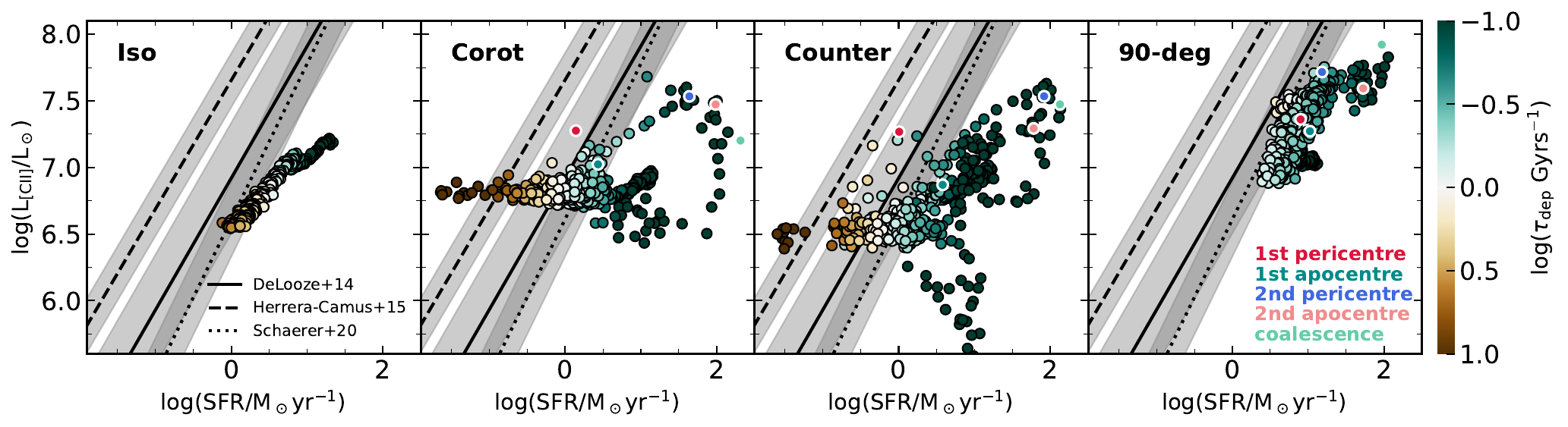}
    \caption{\lcii-SFR relation for the tested scenarios coloured-coded by the galaxy's depletion time. Markers and observed relations are described in Fig.~\ref{fig:lcii_sfr}.}
    \label{fig:lcii_sfr_tdep}
\end{figure*}

To assess the state of our galaxies as they evolve in the \lcii-SFR plane, in Fig.~\ref{fig:lcii_sfr_tdep} we study its dependence on the depletion time. This is defined as the time that the galaxy will require to consume all its molecular gas at its current SFR, i.e. \tdep = $\rm M_\hh/SFR$. 
Considering a normal star-forming galaxy with a \tdep = 1 Gyr, we show in blue colours the starburst phases and in brown the quiescent phases of the evolving galaxies in each simulation.

\begin{figure*}
    \centering
    \includegraphics[width=\linewidth]{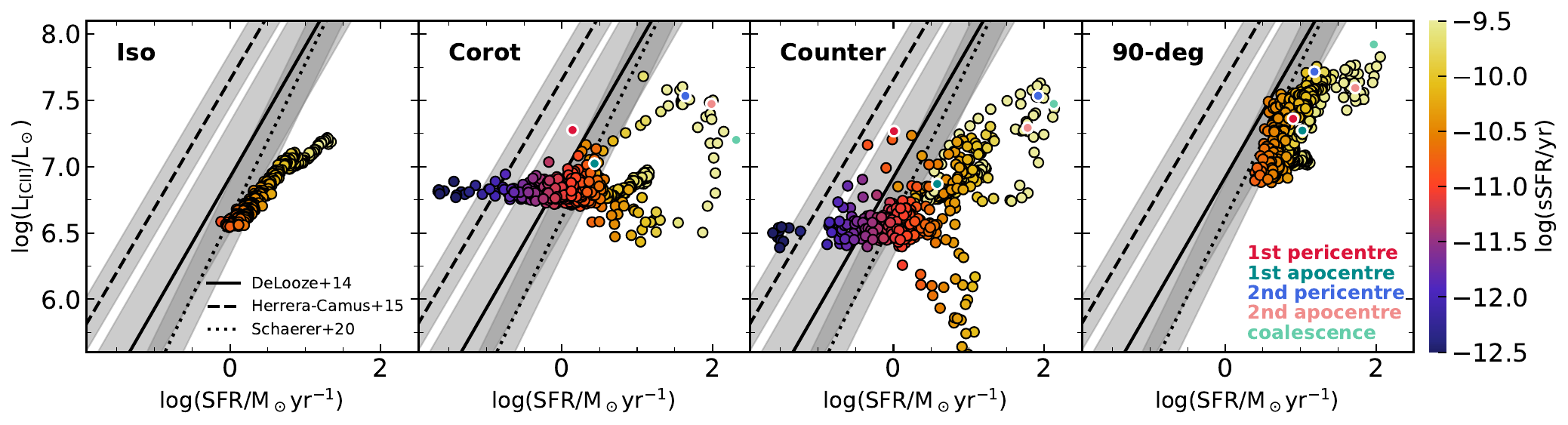}
    \caption{\lcii-SFR relation for the tested scenarios coloured-coded by the galaxy's sSFR. Markers and observed relations are described in Fig.~\ref{fig:lcii_sfr}.}
    \label{fig:lcii_sfr_ssfr}
\end{figure*}

The merger cases cross the observed relations when the galaxies are normal star-forming galaxies, and deviate from the relation when the galaxies are either in a starburst phase with \tdep $\sim$ 0.1 Gyr or in a quiescent phase with \tdep $\sim$ 10 Gyr \citep{Kennicutt1998}. This might indicate that the \lcii-SFR relation does not hold as well in galaxies in extreme phases of their evolution.
Similarly, in Fig.~\ref{fig:lcii_sfr_ssfr} we show the \lcii-SFR plane coloured by the specific star formation rate, sSFR = SFR/\mstar.
Systems that deviate from the relation are either very efficient or very inefficient in transforming gas into stars, implying that even if there is gas available (i.e. short depletion times), it is not in conditions to form stars.

This indicates a more complex evolution during mergers, which affects the thermodynamical properties of the gas in conditions to emit in \cii, and suggests that interactions and/or mergers are efficient mechanisms behind the ionisation of the emitting gas in our simulations. For more detail, see appendix~\ref{ap:bridge_cii}. Our estimations for the simulated cases are in agreement with observed relations \citep{Zanella2018,Madden2020,Heintz2021,Vizgan2022}. In the following section, we explore how \cii behaves as a global cold gas tracer.

\section{\cii as a global gas tracer}\label{sec:tracers}

\begin{figure*}
    \centering
    \includegraphics[width=\linewidth]{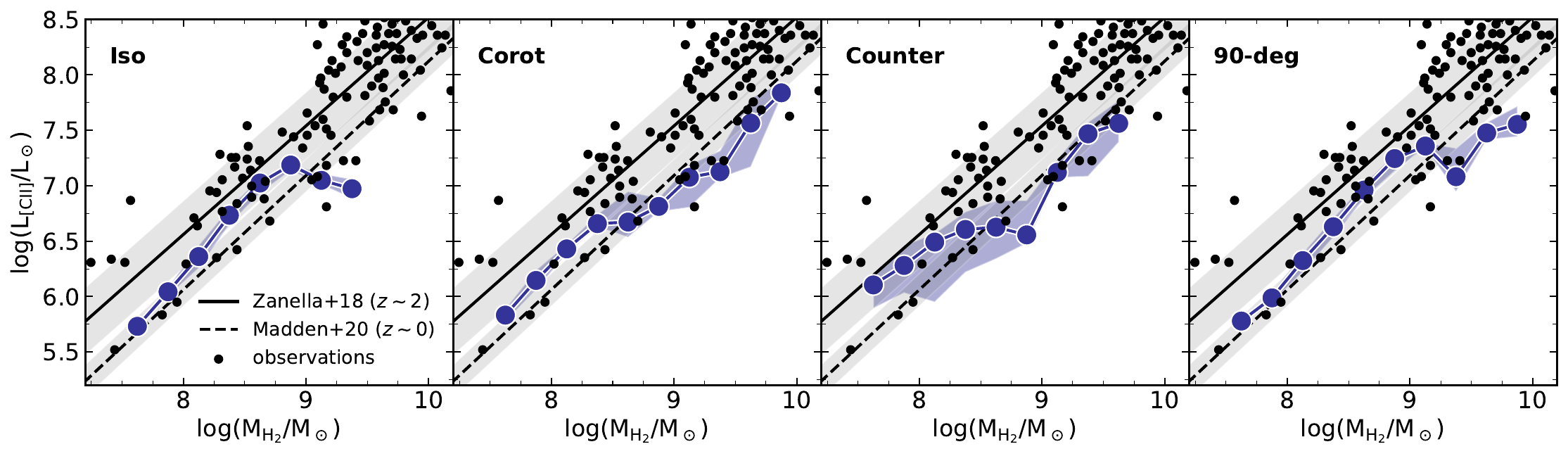}
    \caption{Total \lcii\ts as a function of the system's molecular gas mass \hh. Each column shows the global trend for the tested scenarios as the median value in mass bins of 0.25\ts dex (blue circles) and the 16-84 percentiles (blue shades). We also show the best-fitting relations for star-forming galaxies at $z\sim2$ by \citet[][solid black line]{Zanella2018} and for dwarf galaxies by \citet[][dashed black line]{Madden2020}. Black circles show observations of nearby dwarf galaxies \citep{Cormier2015,Madden2020}, main-sequence and starburst galaxies \citep{Stacey1991,Diaz-santos2013,Diaz-santos2017,Magdis2014,Accurso2017,Contursi2017,Hughes2017}, and high-redshift star-forming galaxies \citep{Ferkinhoff2014,Huynh2014,Capak2015,Gullberg2015,Schaerer2015,Zanella2018,Kaasinen2024}.}
    \label{fig:lcii_hmol}
\end{figure*}

As described in Sect.~\ref{sec:introduction}, \lcii\ts traces the cold gas with temperatures under $\rm T < 10^4 \ts K$. Thus, we analyse how the \cii emission behaves as a gas tracer for \hh, \hi, and \hii.
We study the relation of the \lcii\ts with the total molecular gas content for the four analysed experiments as shown in Fig.~\ref{fig:lcii_hmol}. It is important to note that in this section we are interested in studying the global trends; therefore, we show the median values (blue circles) in mass bins of 0.25\ts dex for each simulation and the 16-84 percentiles (blue shades).

In addition, observational relations, $\log(\lcii/\lsun) = -1.28 + 0.98 \times \log(\rm M_{\hh}/ \msun)$ by \citet[][solid black line]{Zanella2018} and $\log(\lcii/\lsun) = -(2.12/0.97) + (1/0.97) \times \log(\rm M_{\hh}/ \msun)$, by \citet[][dashed black line]{Madden2020}, with the reported scatter, are shown. Observations (black circles) of nearby dwarf galaxies \citep{Cormier2015,Madden2020}, main-sequence and starburst galaxies \citep{Stacey1991,Diaz-santos2013,Diaz-santos2017,Magdis2014,Accurso2017,Contursi2017,Hughes2017}, and high-redshift star-forming galaxies \citep{Ferkinhoff2014,Huynh2014,Capak2015,Gullberg2015,Schaerer2015,Zanella2018,Kaasinen2024} are also displayed.

Our simulations follow the observed relations with \hh, showing mean deviations of 0.6\ts dex, 0.6\ts dex, 0.5\ts dex, and 0.3\ts dex at a \hh\ts mass of $\log(\rm M_\hh/\msun)\sim 9$ with respect to the relation by \citet{Zanella2018} for the \iso, \corot, \counter, and \perpendicular\ts simulations, respectively.
Similarly, a deviation of 0.2\ts dex, 0.1\ts dex, 0.1\ts dex, and -0.1\ts dex, at the same mass, with respect to the relation by \citet{Madden2020}, is estimated for the \iso, \corot, \counter, and \perpendicular\ts simulations, respectively.

In all the scenarios tested, the \lcii\ts correlates strongly with the total molecular mass. Therefore, \lcii\ts proves to be a robust tracer of the total molecular content in our simulated galaxies, even in different stages of their evolution. In fact, we performed a Spearman test and found a correlation coefficient of 0.59 (p < 0.00). We estimated the best-fitting relation for the whole sample of galaxies, including all the simulations and all snapshots, as $\log(\lcii/\lsun) = (1.11\pm0.16) + (0.63\pm0.02) \times \log(\rm M_{\hh}/ \msun)$, in very good agreement with the reported observational slopes.

\begin{figure*}
    \centering
    \includegraphics[width=\linewidth]{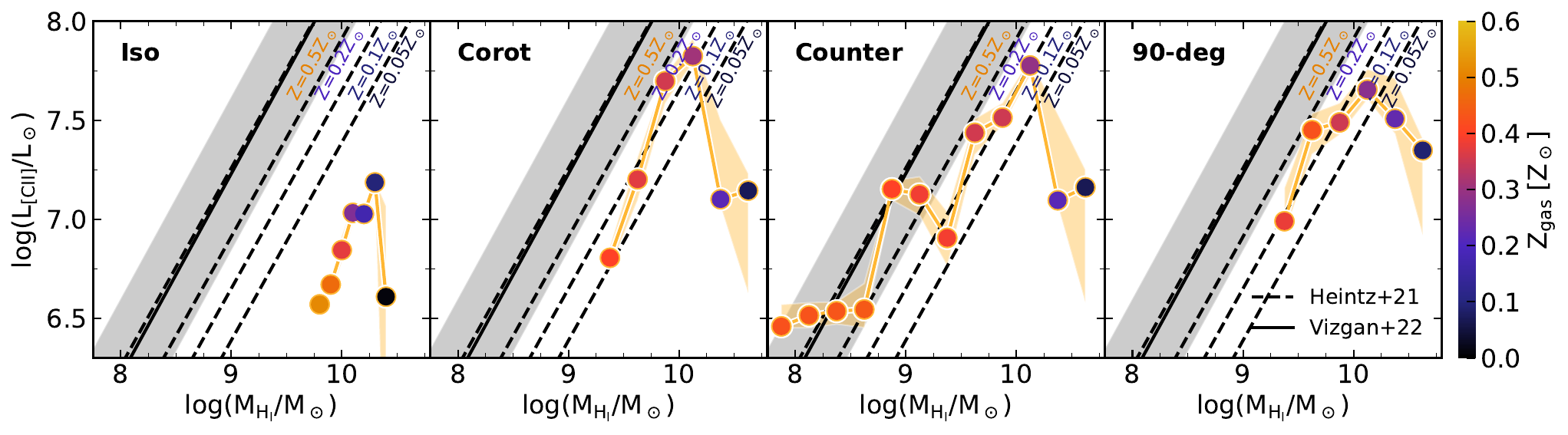}
    \caption{Total \lcii\ts as a function of the system's atomic hydrogen gas mass \hi. Each column shows the global trend for the tested scenarios as the median value in mass bins of 0.1\ts dex for the \iso\ts simulation (first column) and of 0.25\ts dex for the merger simulations (three right columns), coloured by the median gas metallicity (circles) and the 16-84 percentiles (orange shades). Best-fitting relations for observations by \citet[][dashed black lines]{Heintz2021} and for simulated star-forming galaxies \citet[][solid black line]{Vizgan2022}.} 
    \label{fig:lcii_hi}
\end{figure*}

For temperatures lower than 10$^4\rm \ts K$, we expect the gas to be dominated by \hi. Therefore, in Fig.~\ref{fig:lcii_hi} we show the total \lcii\ts as a function of the total \hi\ts gas mass as the median values in mass bins of 0.1\ts dex for the \iso\ts simulations, and of 0.25\ts dex for the merger simulations\footnote{We used different bin size for the \iso\ts and merger simulations to have a fair number of bins in each case. For the \iso\ts case, the evolution of the \hi\ts mass varies in a smaller range than for the mergers; therefore, we have used a smaller bin size.}. In addition, the best-fitting observational relations, $\log(\lcii/\lsun) = (-0.87 \pm 0.09) \times \log(\rm Z/Z_\odot) + (1.48 \pm 0.12) + \log(\rm M_{\hi}/ \msun)$, by \citet[][dashed black lines]{Heintz2021} and for simulated star-forming galaxies, $\log(\lcii/\lsun) = 1.02 \times \log(\rm M_{\hh}/ \msun) -1.95$ by \citet[][dashed black line]{Vizgan2022}, are shown. 
The slopes traced by our simulations are comparable to previous works, and the difference in the zero point of 0.5\ts dex can be attributed to different factors, such as that the \lcii\ts can be primarily dominated by \hh\ts or \hii. We note that the largest difference from the observed relations is detected for the isolated case. The three merger events are in global agreement with observations, suggesting that at the same $\rm M_\hi$ the emitted \lcii\ts is the expected. As can be seen in Fig.~\ref{fig:lcii_hi}, the simulated \lcii\ts positively correlates with the total atomic gas, and a Spearman test estimates a correlation coefficient of 0.63 (p < 0.001). 

Given that the relation by \cite{Heintz2021} considers a dependence on metallicity so that at a given $\rm M_\hi$, a higher metallicity implies larger \lcii, we coloured the circles by the median gas metallicity of each bin, with metallicity defined as the ratio of metals in the gas phase to the total gas mass, $\rm Z_{gas}=M_{Z}/M_{gas}$. 

The simulated \lcii\ts does have a dependence on gas metallicity in some cases, where for a given mass of \hi, higher-metallicity gas has higher \cii luminosities in agreement with \citet{Heintz2021}. More enriched gas tends to have higher luminosities, due to the larger availability of carbon in the medium, and also of \hi, since the higher formation of dust prevents \hi\ts from photoionising.

In the case of the \iso\ts experiment, the simulated galaxy determines a trend with a slope similar to observations, albeit displaced to a higher \hi\ts mass. This could be due to the fact that we can account for the whole HI mass in the simulations, while observations could be missing part of it. On the other hand, this suggests that the \iso\ts experiment is, indeed, too simple in order to reproduce observational trends, and physical mechanisms that strongly disturb the ISM, such as mergers and interactions, are needed in order to generate the diversity of thermodynamical properties present in real galaxies.

We find for a \hi\ts mass of $\log(\rm M_\hi/\msun)\sim$10 a mean deviation with respect to the relation by \citet{Heintz2021} of 2.8\ts dex, 1.9\ts dex, 1.9\ts dex, and 1.9\ts dex for the \iso, \corot, \counter\ts, and \perpendicular\ts simulations, respectively.  
Our estimations are in better agreement with \citet{Heintz2021}, for which we have a mean deviation for a \hi\ts mass of $\log(\rm M_\hi/\msun)\sim$10 of 1.4\ts dex, 0.6\ts dex, 0.6\ts dex, and 0.8\ts dex for the \iso, \corot, \counter\ts, and \perpendicular\ts simulations, respectively. Mergers are chaotic events that impact the chemical evolutionary history of the interacting galaxies, in particular the gas phase \citep[Jara-Ferreira in prep,][]{Perez2011}. Therefore, as expected, the relation of \cii and \hi\ts with the gas metallicity exhibits a more complex behaviour. We estimate the best-fitting relation for the galaxy sample, across their evolution and including all the simulations, as $\log(\lcii/\lsun) = (3.78 \pm 0.11) + (0.32 \pm 0.01) \times \log(\rm M_{\hi}/ \msun) + (-0.18 \pm 0.04) \times \log(\rm Z/Z_\odot)$. In all merger experiments, the simulated galaxies move across the observed regions.

\begin{figure*}
    \centering
    \includegraphics[width=\linewidth]{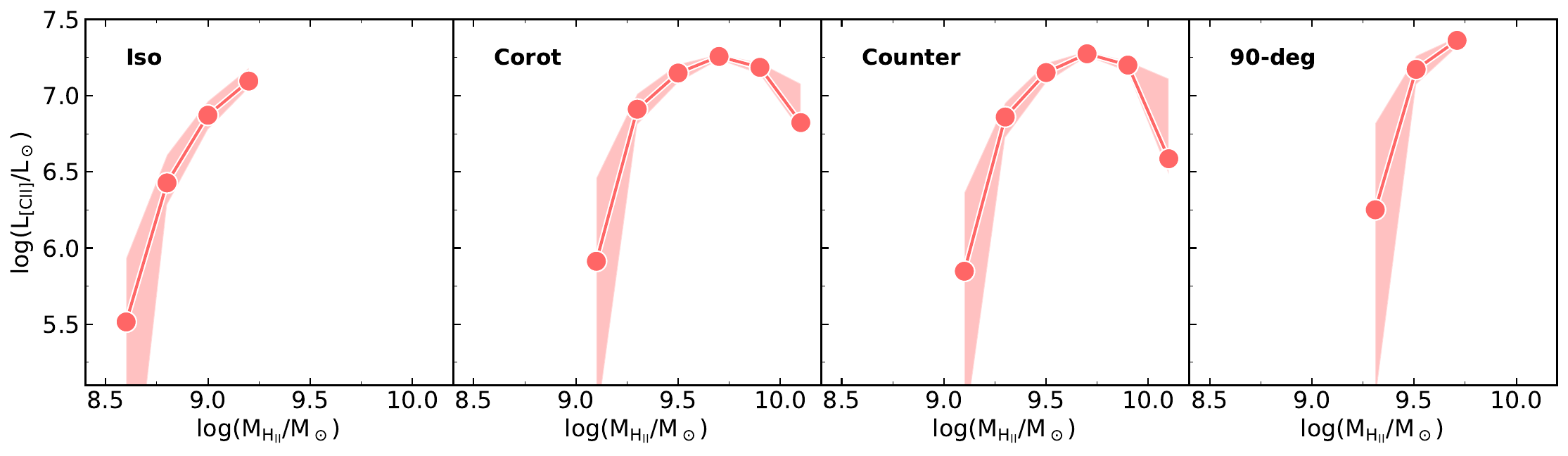}
    \caption{Total \lcii\ts as a function of the system's ionised hydrogen gas mass \hii. Each column shows the global trend for the tested scenarios as the median value in mass bins of 0.2\ts dex (red circles), and the 16-84 percentiles (red shades).}
    \label{fig:lcii_hii}
\end{figure*}

Additionally, we studied the correlation of \lcii\ts with the ionised gas, \hii, as we consider that collisions with \e\ts are implicitly tracing collisions with \hii. In Fig.~\ref{fig:lcii_hii} we show the total \lcii\ts as a function of the total \hii\ts gas mass, for each experiment, as the median values in mass bins of 0.2\ts dex (red circles). We note that mergers are displayed towards higher masses because the two galaxy members are added together in the case of the merger experiments.
For all of our experiments, we find a positive correlation of the \lcii\ts with the total ionised gas mass. However, we notice that this increasing trend flattens out for $\rm M_{\hii} \sim 10^{9.5}\ts \msun$ and falls for $\rm M_{\hii} > 10^{10} \ts \msun$. Currently, there are no available observations of \lcii from \hii\ts gas to compare our results with, but we include our estimations for future reference.

 As was mentioned before, for higher \hii\ts content in the simulated systems, the \lcii\ts saturates. This might indicate that at higher masses of \hii\ts the emission may be driven more predominantly by other gas phases, such as \hh\ts or \hi. Therefore, we can estimate the amount of \cii emission produced by collision with each collider by rewriting Eq.~\ref{eq:lambda} as
\begin{equation}
\begin{aligned}
        L_{\cii,p} =& L_{\cii,p,\hh} + L_{\cii,p,\hi} + L_{\cii,p,\e} \\ 
     =& \frac{n_{\cplus} (n_{\hh}\gamma^{\hh}_{12}+n_{\hi}\gamma^{\hi}_{12}+n_{\e}\gamma^{\e}_{12})}{C}A_{21}\Delta E_{21}
\end{aligned}
,\end{equation}
where 
\begin{equation}
\begin{aligned}
     L_{\cii,p,\hh} =& \frac{n_{\cplus} n_{\hh}\gamma^{\hh}_{12}A_{21}\Delta E_{21} }{C},\\
     L_{\cii,p,\hi} =& \frac{n_{\cplus} n_{\hi}\gamma^{\hi}_{12}A_{21}\Delta E_{21} }{C},\\ 
     L_{\cii,p,\e} =& \frac{n_{\cplus} n_{\e}\gamma^{\e}_{12}A_{21}\Delta E_{21} }{C}
\end{aligned}
,\end{equation}
    with $C = n_{\hh}(\gamma^{\hh}_{12}+\gamma^{\hh}_{21})+n_{\hi}(\gamma^{\hi}_{12}+\gamma^{\hi}_{21})+n_{\e}(\gamma^{\e}_{12}+\gamma^{\e}_{21})+A_{21}$. 

\begin{figure*}
    \centering
    \includegraphics[width=\linewidth]{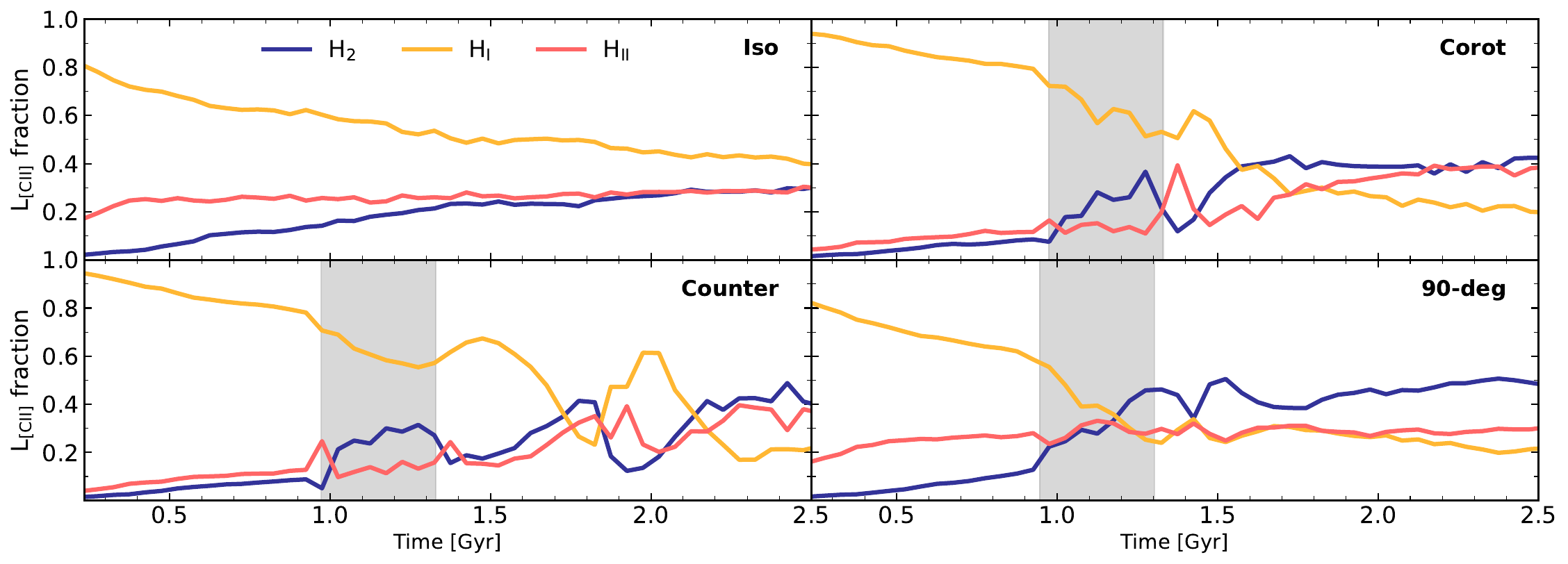}
    \caption{\lcii\ts fraction traced by \hh (blue lines), \hi\ts (orange lines), and \hii\ts (red lines) evolution with time for the emitting gas of the system. Each panel shows the evolution of the trends as the median value in a time bin of 0.05 Gyr for each simulation. The grey band shows the duration of the merger, from the first pericentre until the coalescence.}
    \label{fig:fraction}
\end{figure*}

In Fig.~\ref{fig:fraction}, we show the time evolution of the \lcii\ts fraction traced by the \hh, \hi, and \hii\ts evolution for each simulation. For clarity, we smoothed the trends by showing the median values in 0.05\ts Gyr time bins and indicated the duration of the merger with grey bands. 

For the \iso\ts experiment, although \lcii\ts remains a robust tracer of \hh, it seems that most of the emission is produced by collisions with \hi. This is expected as the gas in temperatures lower than $10^4\rm \ts K$ is in its majority in the form of \hi\ts. Our findings agree with observational trends \citep{Pineda2014,Croxall2017,Cormier2019,Ikeda2025}, in which the contribution of \hi\ts is $\sim$ 40 percent. It should be noted that we have a smaller contribution of \hi\ts with respect to \citet{Casavecchia2025} of $\sim$ 30 percent. We speculate that we are missing part of the \hi\ts contribution given by the CGM gas, since our simulations are not in a cosmological framework and do not contain a CGM or intergalactic medium (IGM). Recently, studies have found an extended \cii emission in the CGM of the Spiderweb galaxy at $z = 2.16$ \citep{DeBreuck2022}, in a quasar host galaxy at $z >$ 6 \citep{Cicone2015}, and in the intracluster medium of the SPT2349-56 protocluster at $z=4.3$ \citep{Harrington2025}.

Globally, the interaction and coalescence of the simulated galaxies produces a strong decrease in the \lcii\ts fraction traced by \hi, by reducing it by $\sim$ 20-30 percent. This increases the contribution of \hh~by $\sim$ 10-20 percent. After the coalescence, most of the \lcii\ts on a global scale is traced by \hh\ts and \hii\ts ($> 40$ percent), whereas for the \iso\ts case, the largest contribution continues to be \hi\ts ($\sim$ 40 percent).

During the first pericentre (beginning of the grey band), a peak in the contribution to the total \lcii\ts by \hii\ts is produced for both \corot\ts and \counter\ts mergers, and a decrease in the contribution by \hh. This is consistent with the decrease in SFR as discussed in Sect.~\ref{sec:model} (see Fig.~\ref{fig:LCII_evolution}). Thus, the emission comes from the ionised that is decoupled from the SF regions. This gas continues to emit \cii forming a bridge during the first apocentre, and remains constantly emitting after the coalescence (see appendix~\ref{ap:bridge_cii}). In a follow-up paper, we explore the spatial extent of this stripped gas and compare how the different configurations of mergers affect this outcome.

Simulations allow us to investigate the distribution of \lcii\ts across different tracers as done above, an analysis that cannot be performed directly with observational data. Therefore,
following \citet{Casavecchia2025}, we studied the contribution to the \lcii\ts dividing the emitting gas by their physical properties into three phases: diffuse (diff, $n_{\rm H} < 10 {\rm \ts cm^{-3}}$), cold dense (CD, $n_{\rm H} > 10 {\rm \ts cm^{-3}}$ and $\rm T < 10^4 \ts K$), and warm dense (WD, $n_{\rm H} > 10 {\rm \ts cm^{-3}}$ and $\rm 10^{4} \ts K  < T < 4 \times 10^4 \ts K$). Additionally, we estimated the fraction of emitting gas that is star forming, i.e. the instantaneous SFR $>$ 0\ts \msun/yr.

\begin{figure*}
    \centering
    \includegraphics[width=\linewidth]{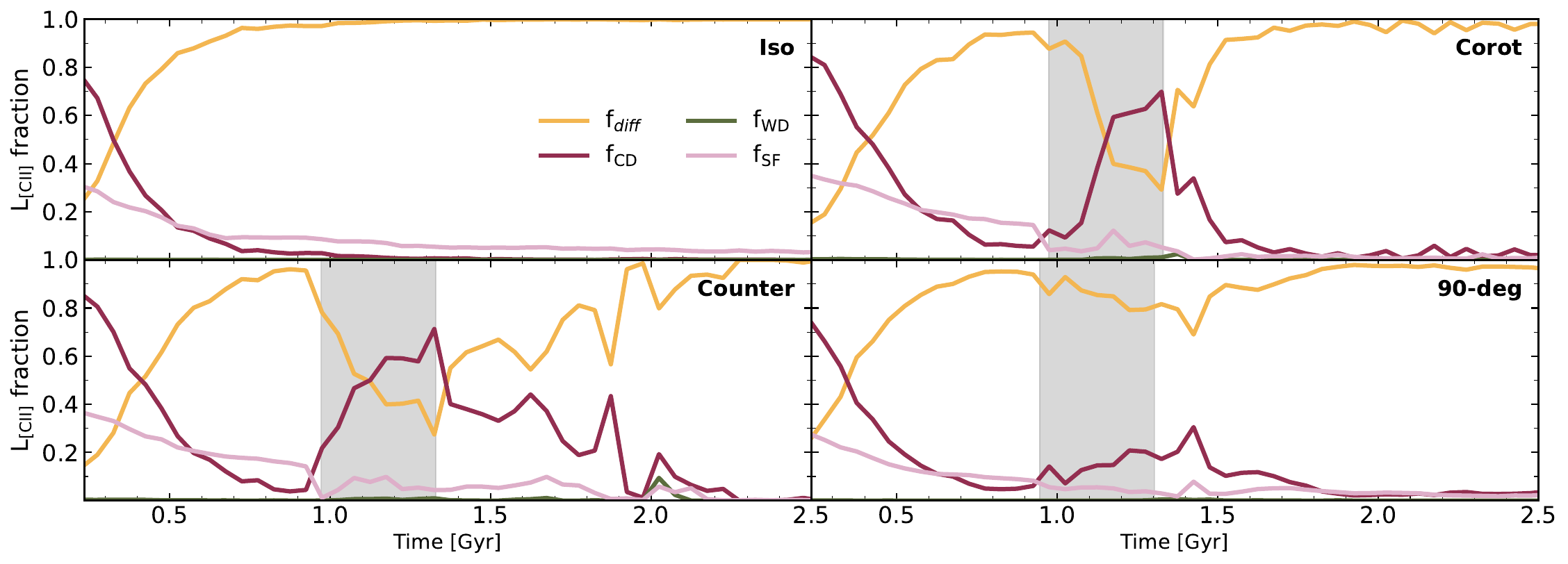}
    \caption{\lcii\ts fraction in the form of diffuse (orange lines), cold dense (burgundy lines), warm dense (green lines), and star-forming (light pink lines) gas evolution with time for the system as a whole. Each panel shows the evolution of the trends as the median value in a time bin of 0.05 Gyr for each simulation. The grey band shows the duration of the merger, from the first pericentre until the coalescence.}
    \label{fig:phases}
\end{figure*}

We show in Fig.~\ref{fig:phases} the evolution of the fraction of the emitting gas in diffuse ($f_{diff}$, orange lines), cold dense ($f_{\rm CD}$, burgundy lines), and warm dense ($f_{\rm WD}$, green lines) states, and the star-forming gas ($f_{\rm SF}$, light pink lines) in each panel for the whole system of the \iso, \corot, \counter, and \perpendicular\ts simulations. As seen in the upper left panel of Fig.~\ref{fig:phases}, as the \iso\ts system evolves with time, the fraction of emitters in the cold dense phase decreases from $\sim$ 80 percent initially, to $<$ 10 percent, whereas the fraction of emitting gas in the diffuse phase rapidly increases, reaching $\geq$ 90 percent. The star-forming gas shows a similar behaviour to the cold dense phase, as expected, since cold and dense gas has the conditions to form stars. 

This behaviour is also present in the mergers just before the interactions start (shaded regions). During the interactions and then coalescence (see upper right and bottom panels), the fraction of emitting gas in the diffuse phase decreases by $\sim$ 60 percent for the \corot\ts and \counter\ts mergers, and by $\sim$ 20 percent for the \perpendicular. During the interaction, the fraction of diffuse gas has a median of $\sim$ 50 percent, in contrast to the $>$ 90 percent of the \iso\ts case at the same time of evolution. This is because of the rapid starburst generated during the interaction, which makes the \cii trace the cold and dense gas more effectively (see Fig.~\ref{fig:fraction}). After the coalescence, the mergers rapidly restore the fraction of emitting gas in the diffuse phase, which accounts for almost its totality. We find no significant contribution to the luminosity of warm and dense gas.

From Fig.~\ref{fig:fraction} and~\ref{fig:phases}, we infer that for the \iso\ts and before the interaction for the mergers, most of the \cii emission comes from \hi\ts that is in cold dense gas. As the SF episodes warm the medium, the \hi\ts becomes diffuse. During the interaction, the rapid starburst brings the emitting gas to cold and dense phase, and thus a larger fraction of \cii comes from \hh\ts gas. Globally, the mergers affect how effectively \hi\ts emits in \cii, making \hh\ts in diffuse state a more predominant contributor to the total \lcii. Fig.~\ref{fig:phases} shows consistently that the peak in \lcii\ts around the first pericentre (beginning of the grey band) corresponds to gas that is being ionised by the interaction and that is decoupled from the SF, since it shows a decrease in the fraction of the star-forming gas contribution on the total \lcii. In the following paper, we study how this ionised gas that was produced during the interaction, and that is decoupled from the SF, is spatially located in the system as we look for extended emission.

\section{Discussion and conclusions} \label{sec:conclusions}

   In this work, our main goal is to understand how \cii behaves as a cold gas tracer in isolated and merging galaxies, and to provide insights into the physics of \cii emission, as well as predictions for future observations. For this purpose, we analysed pre-prepared simulations by Sillero in preparation run with \pgadgetk\ \citep{Sillero2021} of galaxy mergers of a pair of MW-mass-size galaxies in three different configurations: coplanar co-rotating (\corot), coplanar counter-rotating (\counter), and perpendicular (\perpendicular), as well as an isolated counterpart (\iso).

    We successfully implemented a semi-analytical model by \citet{Casavecchia2025} to estimate \lcii, and studied the distribution of the \cii emission, and \cplus, \hh, \hi\ts, and \hii\ts abundances, as well as the gas properties. We summarise our main results and conclusions as follows:

\begin{enumerate}
    \item As seen in Fig.~\ref{fig:LCII_evolution}, we find that the total \lcii\ts increases with strong star formation episodes. In particular, for the mergers, the \lcii\ts rapidly increases at the pericentres of the interaction, and remains constant long after the coalescence of the galaxy pair.
    \item Our simulations follow the expected \lcii-SFR relation as shown in Fig.~\ref{fig:lcii_sfr}. The isolated galaxy is in good agreement with the observed relations by \citet{HerreraCamus2015} and \citet{Schaerer2020}. Even though mergers show a more chaotic behaviour, the galaxies follow the positive correlation with the SFR, indicating that both the star formation and \lcii\ts are linked. After the coalescence, the resulting systems are located within the observed range, settling into a constant \lcii\ts ($\log{(\lcii/\rm L_\odot)}\sim$ 6.5-7); that is, for a diversity of SFRs, a determined \lcii\ts is observed. This points to a decoupling between the SFR and the \lcii, in which additional physical processes might also be contributing to the \cii emission on a global scale, in our simulations.
    We find that the \lcii-SFR relation correlates with the gas oxygen abundance, where for a given SFR, galaxies with higher \lcii\ts have higher levels of enrichment of their gas (see Fig.~\ref{fig:lcii_sfr_OH}). The simulated galaxies deviate from the \lcii-SFR relation when they are either starbursts ($\tdep \sim 0.1$\ts Gyr) or quiescent ($\tdep \sim 10$\ts Gyr), suggesting that the \lcii-SFR might not hold in extreme phases of the evolution, at least within the framework of our subgrid physics model. Similarly, Fig.~\ref{fig:lcii_sfr_ssfr} shows that our simulated galaxies with very efficient or very inefficient star formation deviate from the relation, independently of the gas content of the system.
    \item We show in Fig.~\ref{fig:lcii_hmol} that \lcii\ts is a robust tracer of molecular gas in our simulated galaxies, even at different stages of their evolution, agreeing with observational relations of \citet{Zanella2018} and \citet{Madden2020}. In our simulations \cii is also a good tracer of neutral hydrogen \hi, and is found to have a weak correlation with the gas metallicity, in agreement with previous works \citep{Delooze2014,Vallini2015,Lagache2018}.
    We find a positive correlation of \lcii\ts with \hii, which flattens at $\rm M_{\hii}\sim10^{9.5}\ts\msun$ until it decreases for $\rm M_{\hii}\geq10^{10}\ts\msun$. Since currently there are no available observations, we report these results to be compared with future observations.
    \item For the merger simulations, the \cii emission on a global scale traces mostly \hh\ts($\sim$ 50 percent), while \hi\ts and \hii\ts account for 30-40 percent of the total luminosity. This is in agreement with observational studies \citep{Pineda2014,Croxall2017,Cormier2019,Ikeda2025}. We find that interactions can decrease the fraction of the emission traced by the diffuse warm \hi\ts gas by $\sim$ 20-30 percent. Globally, our simulated mergers show that \hh\ts in diffuse state is a more predominant contributor to the total \lcii\ts compared to the \iso\ts galaxy.

   Our finding suggest that \cii emission involves a complex interplay of different physical properties of the gas phase. We have explored how \cii behaves as a cold gas tracer on a global scale during interactions and mergers of controlled simulated systems, providing new insights. However, it remains to be explored the \cii behaviour on resolved scales. In a following paper, we visit this idea by doing a spatially resolved study and looking for extended \cii emission in galaxy mergers.
    
\end{enumerate}
\begin{acknowledgements}
 We thank the anonymous referee for their insightful comments which helped improve the clarity of this manuscript. We thank the EvolGal4D team and the High-redshift Galaxies team for the interesting discussions throughout this work. The authors gratefully thank Manuel Aravena for the insightful discussions that helped define the scope of this work.
 VPM acknowledges funding by ANID (Beca Magíster Nacional, Folio 22241063). PBT acknowledges partial funding by Fondecyt-ANID 1240465/2024. JGL gratefully acknowledges support from ANID MILENIO NCN2024\_112. VPM and JGL acknowledge support by FONDECYT grant No. 1252054. PBT and JGL acknowledge support from ANID BASAL project FB210003. ES acknowledges funding by Fondecyt-ANID Postdoctoral 2024 Project N°3240644. We acknowledge support from the European Research Executive Agency HORIZON-MSCA-2021-SE-01 Research and Innovation programme under the Marie Skłodowska-Curie grant agreement number 101086388 (LACEGAL). This project used the Ladgerda Cluster (Fondecyt 1200703/2020 hosted at the Institute for Astrophysics, Chile), Geryon 3 Cluster (ANID Basal Project FB210003) and the Barcelona Supercomputer Center (Spain).
      
\end{acknowledgements}

\bibliographystyle{aa} 
\bibliography{biblio.bib}

\begin{appendix}
\section{The formation of a \cii bridge}\label{ap:bridge_cii}
\begin{figure}
    \centering
    \includegraphics[width=\linewidth]{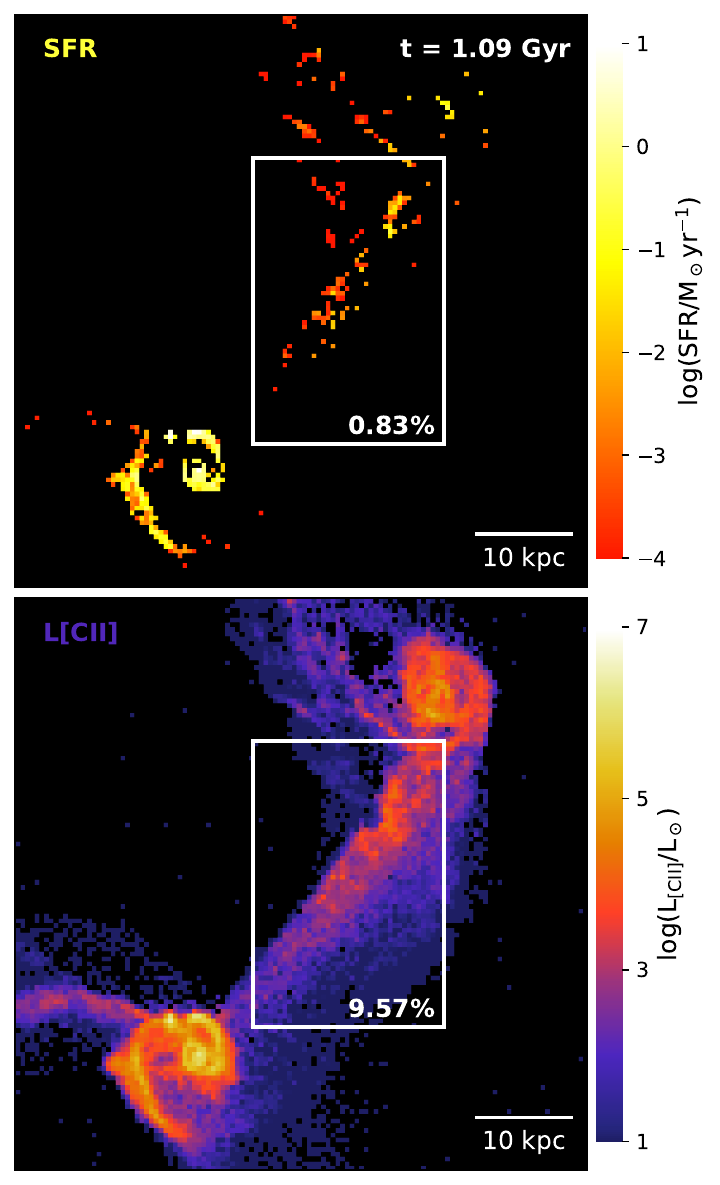}
    \caption{Projection maps of instantaneous SFR (upper panel) and \lcii (bottom panel) for the \counter\ts merger at the moment of 1st apocentre. White rectangle denotes a 20\ts kpc$\times$25\ts kpc region enclosing the gas bridge between the galaxy pair. Inset labels show the percentage of the total instantaneous SFR (upper panel) and \lcii (bottom panel) within this region.}
    \label{fig:bridge}
\end{figure}

We analyse the formation of a bridge of gas emitting in \cii for the \counter\ts simulation during the 1st apocentre as an example in Fig.~\ref{fig:bridge}. In the upper panel, a projection map of the gas instantaneous SFR is shown, whereas the bottom panel shows a projection map of \lcii. Both maps were obtained by doing a projection in the z axis with a pixel size of 0.5\ts kpc.
We selected a rectangle of 20\ts kpc$\times$25\ts kpc, enclosing the \cii bridge between the galaxy pair. Inset labels show the percentage of the total instantaneous SFR (upper panel) and of total \lcii\ts (bottom panel) within this region at the first apocentre.

Whereas only 0.83 percent of the total instantaneous SFR is located in the bridge, the \cii emission accounts for almost $\sim$ 10 percent. The predominance of \cii over the SFR in the gas bridge is prevalent during the interaction, pointing to a decoupling between the emission of \cii and the presence of young and hot recently formed stars. This stripped gas continues to emit long after the coalescence of the galaxy pair.

\section{Adopted constants}
In Table.~\ref{table:constants} we present the constants used to estimate the \lcii\ts and the respective reference, following the model by \citet{Casavecchia2025}.
\begin{table}[h]
\caption{Adopted constants for the \cii model implemented.}
\label{table:constants}     
\centering                      
\begin{tabular}{c c c}   
\hline\hline            
\noalign{\vskip 3pt}
Constant & Value &  Reference \\[3pt]
\hline
\noalign{\vskip 4pt}
$\gamma^{\e}_{21}$ & $2.8\times10^{-7} T^{-0.5}_{100}{\rm cm^{3}s^{-1}}$ & {(3)}\\[4pt]
$\gamma^{\hi}_{21}$ & $8\times10^{-10} T^{0.07}_{100}{\rm cm^{3}s^{-1}}$ & {(3)}\\[4pt]
$\gamma^{\hh}_{21}$ & $3.8\times10^{-10} T^{0.14}_{100}{\rm cm^{3}s^{-1}}$ & (2) \\[4pt]
$A_{21}$ & $2.4\times10^{-6} {\rm s^{-1}}$ & (1) \\[4pt]
$\Delta E_{21}$ & $1.259\times10^{-14} {\rm erg}$ & (4) \\[4pt]
T$_{\rm exc}$=$E_{21}$/k & 91.2 K& {(3)}\\[4pt]
\hline    
\end{tabular}
\tablefoot{Columns from left to right contain the constant used, the value, and the reference. References: (1) \citet{Draine2011}, (2) \citet{Goldsmith2012}, (3) \citet{Hollenbach&McKee1989}, and (4) \citet{Santoro2006}.}
\end{table}
\end{appendix}

\label{LastPage} 
\end{document}